# Conservation or deterioration in heritage sites? Estimating willingness to pay for preservation

**Ardeshiri A.[1*], Etminani-Ghasrodashti R.[2], Hossein Rashidi T.[1], Ardeshiri M.[3,4], Willis K.[5]**

1- Research Centre for Integrated Transport Innovation (rCITI), School of Civil and Environmental Engineering, University of New South Wales (UNSW) Sydney, NSW 2052 Australia.
2- Department of Urban Planning, Science and Research Branch, Islamic Azad University, Tehran, Iran.
3- Department of Urban Planning, Shiraz University, Shiraz, Iran.
4- Department of Urban Planning, Pishtazan Higher Education Institute, Shiraz, Iran.
5- School of Architecture, Planning and Landscape, University of Newcastle upon Tyne, UK

\* Corresponding author – email: A.Ardeshiri@unsw.edu.au

Declarations of interest: *none*

## ABSTRACT:

A significant part of the United Nation's World Heritage Sites (WHSs) is located in developing countries. These sites attract an increasing number of tourist and income to these countries. Unfortunately, many of these WHSs are in a poor condition due to climatic and environmental impacts; war and tourism pressure, requiring the urgent need for restoration and preservation (Tuan & Navrud, 2007).

In this study, we characterise residents from Shiraz city (visitors' and non-visitors') willingness to invest in the management of the heritage sites through models for the preservation of heritage and development of tourism as a local resource. The research looks at different categories of heritage sites within Shiraz city, Iran. The measurement instrument is a stated preference referendum task administered state-wide to a sample of 489 respondents, with the payment mechanism defined as a purpose-specific incremental levy of a fixed amount over a set period of years.

A Latent Class Binary Logit model, using parametric constraints is used innovatively to deal with any strategic voting such as "Yea-sayers" and "Nay-sayers", as well as revealing the latent heterogeneity among sample members. Results indicate that almost 14% of the sampled population is unwilling to be levied any amount ("Nay-sayers") to preserve any heritage sites. Not recognizing the presence of nay-sayers in the data or recognizing them but eliminating them from the estimation will result in biased Willingness to Pay (WTP) results and, consequently, biased policy propositions by authorities. Moreover, it is found that the type of heritage site is a driver of WTP.

The results from this study provide insights into the WTP of heritage site visitors and non-visitors with respect to avoiding the impacts of future erosion and destruction and contributing to heritage management and maintenance policies.

# 1. INTRODUCTION

Expansion of cities, population growth, changes in the life style, improving the quality of life are examples of the inevitable movements which almost all cities around the world are experiencing. Although this new progress has brought many opportunities and benefits in terms of development and socio-economic welfare, it has also triggered some challenges for sustainable urban planning, particularly in terms of conservation and preservation of cultural heritage (Seyedashrafi, Ravankhah, Weidner, & Schmidt, 2017).

Heritage is one of the most rapidly expanding tourism segments regarding visitor numbers globally and is a major attraction for cities and regions. Heritage sites and buildings can have a very positive influence on many aspects of the way a community develops. If heritage is managed appropriately, it can become instrumental in developing intercultural dialogue, enhancing social inclusion, shaping identity of a territory, providing social cohesion, improving quality of the environment. Moreover, from the economic side, it can create jobs, prompt tourism development and enhance investment climate (Dümcke & Gnedovsky, 2013). particularly any investment in heritage can generate return in a form of economic growth and social benefits. The historic environment has been accepted as a source of benefit to local economies, predominantly through tourism (Mowforth & Munt, 2015). An attractive heritage, supports and contributes in enticing external investment as well as maintaining existing businesses of all types. The heritage environments are also seen as an excellent resource for education purposes for people of all ages (Talboys, 2016). Learning about the history of a place is a good way of bringing communities together through a common awareness of the unique cultural identity heritage places give to an area. Moreover, heritage adds character and distinctiveness to an area and is a fundamental in creating a 'sense of place' for a community (Graham, Ashworth, & Tunbridge, 2016; Massey, 2012).

While development and use provides' many opportunities for the community, it may also serve as a threat in forms of the potential degradation of a heritage, thus depriving a community of such resources and the benefits of tourism. The connection between heritage and tourism is commonly characterized by contradictions and conflicts whereby conservationists perceive heritage tourism as compromising conservation goals for profit (Nuryanti, 1996). In order to minimize these threats and ensure that the heritage is passed onto the next generation, no different than the current status, there is a need for sustainable dialogue, cooperation, and collaboration between the community and heritage management (Aas, Ladkin, & Fletcher, 2005). Heritage preservation demands an ongoing need for management (McKercher, McKercher, & Du Cros, 2002). Total protection and severe restriction of visitation is a widely used management approach (Grimwade & Carter, 2000). However, it can be argued that the approach contradicts with the value in preserving such places for the future if people are not permitted to enjoy them today! Often heritage sites are subjected to well-managed restoration and conservation actions, but the critical component of presentation and interpretation to visitors, both tourists and local residents is absent. Without suitable appreciation and presentation of what is being preserved, cultural heritage sites potentially become meaningless, and understanding of human history is lost.

Different heritage management strategies are likely to be relevant at different heritage sites dependent on community preferences for the configuration of the heritage site, including the appropriate mix of built versus natural assets, into the future. Irrespective of the management option that is ultimately selected in a given setting, securing heritage sites for future use requires significant investment in long-term planning and management. To ensure effective and efficient future management, it is necessary to a) enhance the financial sustainability to ensure the level of funding allocated to heritage preservation management is sufficient for management costs, b) ensure that management is in line with community values and preferences for future heritage configurations and c) move away from ad-hoc protection and repair works towards a more strategic management approach that prioritises the maintenance and protection of natural and/or built assets at key priority locations. Multiple questions then arise including: are citizens willing to invest in the maintenance of their heritage sited? how does management account for affected parties (like homeowners) as well as other stakeholders, like those who use heritage sites for recreation, or those who place a high value on the preservation of the sites? which heritage site(s) should receive higher funding priority?

Addressing these questions requires a better understanding of the full suite of economic values the population holds for heritage assets. In this context, it is information relating to the non-use values of heritage sites that is currently most lacking. These non-use values encompass existence value –the value associated with knowing that biodiversity and other environmental values continue to exist (Perace & Moran 2013), bequest value – "a willingness to pay to preserve the environment for the benefit of one's descendants" (Turner et al. 1994) or for future generations (Pearce & Moran 2013), and option value – a value people place on potential future use of an environmental site or resource (Stevens et al. 1991). These are best quantified using non-market stated preference techniques like Contingent Valuation and Choice Modelling (see Methods).

This study seeks to quantify non-use values for a specific heritage asset types held by households in Shiraz in southwest of Iran. We have designed this study so that it will address a number of the management challenges identified above. First, the payment mechanism employed is a targeted incremental annual levy. This can provide managers with an estimate of residents' willingness to pay for management that preserves heritage sites, as well as a realistic vehicle through which sustainable financing of heritage management might be achieved. Second, we employ a repeated, hypothetical referendum task to compare willingness to pay (WTP) among three heritage site categories (Gardens, Historical places and religious sites) and in response to the travel distance between a specific heritage site location and the respondent's home. These elements of our survey design address the question of how investment should be prioritised amongst a range of heritage site available in Shiraz and Pasargadae counties that are exposed to erosion processes.

From an applied point of view, this study contributes to the current literature by providing significant empirical findings that heritage managers can benefit from in their decision making about how to sustainably and efficiently finance heritage management int the future. More importantly, from a methodological perspective, this study uses an innovative latent class model to infer both preference heterogeneity and to identify and deal with strategic (or protest) voting in the form of "nay-" as well as "yea-saying" during estimation. We present an alternative approach for dealing with nay- and yea-saying; many studies exclude these groups from overall estimates of WTP, which we argue can lead to a serious under- or over-estimation of value.

The remainder of the article proceeds as follows. First, we provide some background information related to the case study. We follow with a description of the method and data used for the study. In the penultimate section, we report results of residents' willingness to pay to preserve heritage sites. We conclude with a discussion of the policy and research implications of our findings.

## 2. BACKGROUND AND CASE STUDY CONTEXT

This study was conducted in Shiraz city in Fars province, Iran. Shiraz as the capital of Fars province is the largest metropolitan area with 1264 $km^2$, and a population of 1.7 million, recognized as the historic-cultural hub in southwest of Iran (Shiraz Municipality, 2015). Fars province has about 288 cultural-historic sites, 125 natural sites and 15 natural-historic sites. More than 90% of cultural-historic sites and all of man-made attractions are located in Shiraz and the immediate suburbs. Furthermore, the Marvdasht-Shiraz corridor is one the most prominent tourism sites of Iran with an international function (Comprehensive Plan of Shiraz Tourism and Branding, 2017).

Based on the Shiraz Revised Detailed Plan, this city is divided to 10 regional municipalities with approximately 80 districts (Shahrokhaneh Consultant, 2014). Shiraz is a historical city and generally has three particular parts including the central business district (CBD) which is mostly recognized as the historical core of the city with a compact built environment, the inner-city suburbs that include areas developed around the CBD, and the outer and exurbs suburbs that have been expanded in northwest, south and southwest of the city in the past 20 years and comprises sprawled neighbourhoods (Etminani-Ghasrodashti et al., 2018). Cultural heritage sites are mainly located in the CBD of the city in particularly region 8 of the Shiraz municipality. The area of the historical centre is approximately 360 hectares, which surrounds about 3 % of the total city. This historical core consists of cultural historic axes which includes monuments, old bazar, gardens and mosques. Some cultural heritage sites are within the inner-city suburbs. In addition to historic centre of the city, there are also reputable ancient-cultural sites throughout the province such as Persepolis, Pasargadae and Necropolis (Naqsh-e-Rostam) that have been listed by the World Heritage Centre (UNESCO) as world heritage sites (see Figure 1). Pasargadae and Persepolis were known as the main governmental and ceremonial capital of the Achaemenid Empire in $6^{th}$ century BCE. Pasargadae is situated in 130 km north of Shiraz and Persepolis is placed in 60 km northeast of Shiraz. Another monument is Naqsh-e-Rostam which is located in 12 km northwest from Pasargadae and is the necropolis of the Achaemenid dynasty.

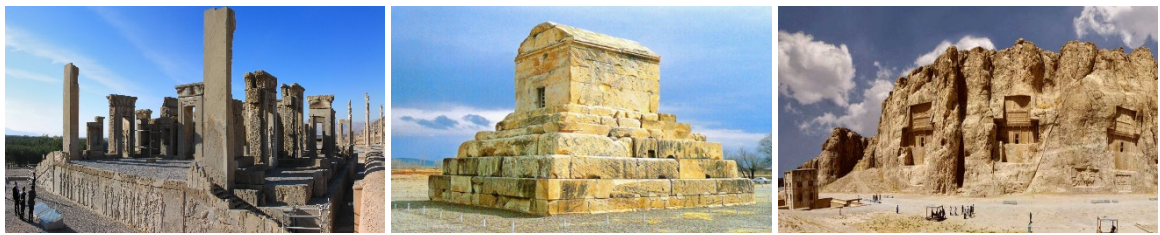

**Figure 1:** examples of world heritage sites in Shiraz; Persepolis, Pasargadae and Necropolis (Naqsh-e-Rostam)- respectively from left to right

In this study, the heritage sites have been categorised into three categories; historic sites, religious sites and gardens. In addition to the historical sites mentioned in Figure 1, other examples of historical sites and their name have been illustrated in Figure 2. Karim-khan Citadel is a complex remained from Zand dynasty and is located in the historic centre of Shiraz. Hafez Mausoleum and Sa`di Mausoleum have memorial structures of reputable Persian poets and erected in the northern edge of Shiraz.

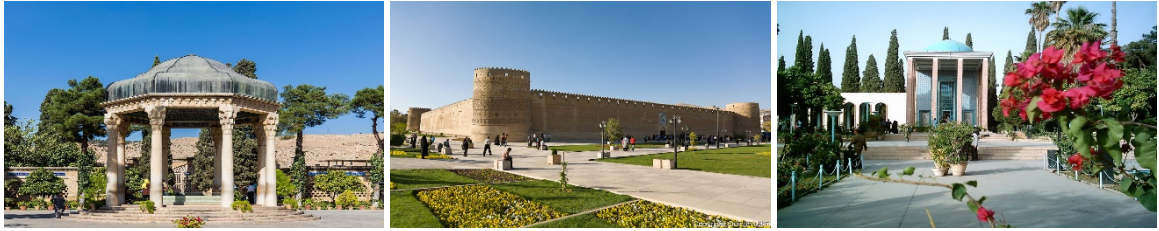

**Figure 2:** examples of historical sites; Hafez Mausoleum, Karim-khan Citadel and Sa`di Mausoleum - respectively from left to right

Figure 3, presents examples of religious sites such as Ali Ebn-e Hamze, Shah-e Cheragh and Vakil Mosque. Shah-e Cheragh is a monumental mosque (14th century), Ali Ebn-e Hamze is a shrine of a person with lineage to the prophet Mohammad (19th century), Vakil Mosque situated to the west of the Vakil Bazaar (18th century).

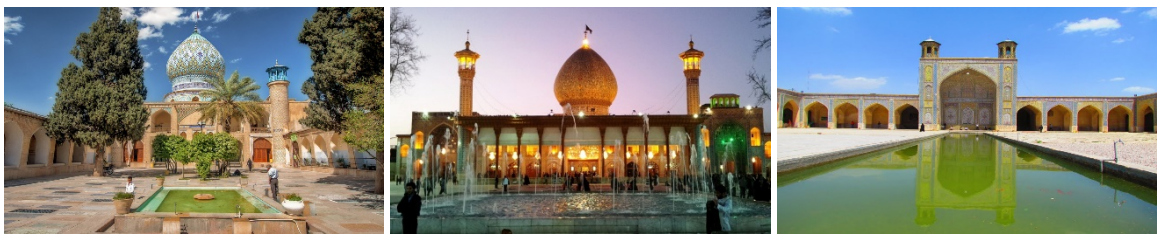

**Figure 3:** examples of religious sites; Ali Ebn-e Hamze, Shah-e Cheragh, Vakil Mosque - respectively from left to right.

Finally, in Figure 4, we have named few Gardens in Shiraz. These gardens are Delgosha Garden, Eram Garden and Jahan-nama Garden - respectively from left to right. Eram garden is a historic Persian garden located at the northern shore of the Khoshk River in the city of Shiraz, Delgosha Garden is located close to the Tomb of Sa`di Mausoleum and Jahan-nama Garden is located near the Hafez Mausoleum.

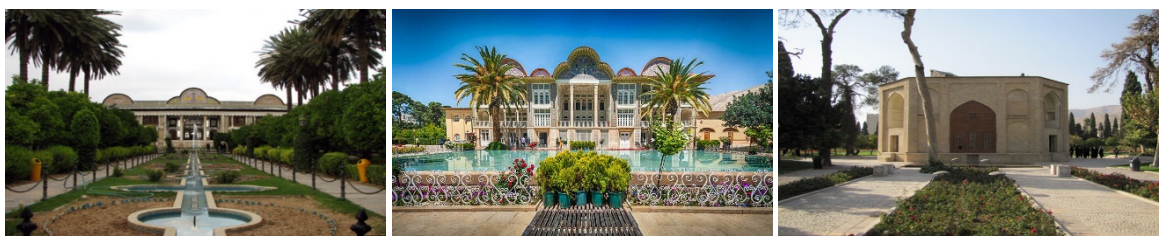

**Figure 4:** examples of gardens; Delgosha Garden, Eram Garden and Jahan-nama Garden - respectively from left to right

According to Comprehensive Plan of Shiraz Tourism and Branding (2017), the visitor rates to heritage sites in shiraz have increased significantly during 2001 to 2016. Based on Tourism Area Life Cycle model (Butler, 1980), Shiraz heritage sites are in different stages of their lifecycle. These stages include: discovery, involvement, development, consolidation and stagnation. For example, Shah-e Cheragh as a religious site is at consolidation stage where the growth of national visitors could result in local residents' dissatisfaction specially by those who have not been participated in the economic advantages of tourism. The visitors of Karim-Khan Citadel have been increased during the past 13 years. However, the proportion of its visitors is dramatically lower than other sites. Hence, this site is still at development stage. Hafez Mausoleum is estimated to be at consolidation stage due to the increasing rate of non-local visitors and probable dissatisfaction of local residents. Controlling the tourism threshold is regarded as the main goal for this site. Sa`di Mausoleum is at development stage. With an average number of 859456 national and international visitors per year, Persepolis is between development and consolidation stages. Although the number of visitors to Pasargadae and Necropolis sites has increased during the past 13 years, the lifecycle of these sites shows that they are passing from involvement stage to development stages and require more investments.

## 3. MATERIALS AND METHODS

### 3.1 THE CHOICE EXPERIMENT

One set of approaches capable of eliciting environmental preference is the use of stated preferences techniques (Ardeshiri, Willis, & Ardeshiri, 2018; Boxall, Adamowicz, Swait, Williams, & Louviere, 1996). Contingent valuation method as part of the stated preferences technique has been widely used in majority of valuation studies of cultural heritage and recently there have also been a few applications of the choice modelling approach (Tuan & Navrud, 2007). Choice modelling has been developed as an alternative type of stated preference technique which is capable of estimating values for changes in resource attribute use where data are not available from markets (J. J. Louviere, D. A. Hensher, & J. D. Swait, 2000). Referendum Choice Experiments (RCE), in the dichotomous choice format, have recently become widely used as a technique for eliciting the value of public goods or non-market resources in applications where the key policy issue is whether to accept an exogenously specific proposal or not (Cameron, 1988; Green, Jacowitz, Kahneman, & McFadden, 1998; Johnston et al., 2017; Rolfe & Bennett, 2009). Respondents are presented with a hypothetical referendum that specifies a good to be supplied and a payment and asked to vote on this referendum (Green et al., 1998). The payment, or *bid*, is varied experimentally to provide a profile of the cumulative distribution function of WTP at the experimental design points. This protocol has gained widespread use in applications to valuation of natural resources and has largely displaced older protocols in which subjects are asked to state an open-ended WTP for a good, or to reveal a WTP range by responses to a sequence of bids or choices from a set of alternatives. Moreover, practitioners have found that responses are influenced by the payment vehicle. This may arise from incentive effects of the 'free-rider' variety, or from the concerns of subjects about distributional implications and 'fairness' (Green et al., 1998).

The concepts in economic theory underlying referendum surveys are preferences characterized in monetary units (*consumer surplus, compensating variation, willingness to pay*), the *Kaldor–Hicks compensation principle* as a criterion for aggregating individual preferences into a social choice rule, and Samuelson's theory of optimal supply of public goods, developed in a stream of literature that has emphasized *incentive-compatible mechanisms* that blunt the 'free-rider' problem (Green, et al., 1998). To be incentive compatible, a referendum on a pure public good needs to be a "*take-it*" or "*leave-it*" offer, where the vote doesn't influence any other offers that may be made to agents and where the payment mechanism is coercive in the sense that each agent can be required to pay independently of how the individual agent voted (Carson & Czajkowski, 2014). Many economists believe that if subjects are adequately economically motivated, the cognitive paradoxes sometimes observed in psychological experiments disappear (Green et al., 1998). Thus, a decision rule should be selected that is realistic and binding on respondents. In many developed countries, political settings direct democracy is practiced exercising majority rule. As a consequence, referenda have been used before to determine the provision of public goods (Green et al., 1998; Mitchell & Carson, 2013). We therefore propose in this research to utilize an RCE approach, focused explicitly on whether Shiraz residents are willing to invest in the management of heritage sites through preservation measures.

## 3.2 EXPERIMENT MATERIALS

Following a literature review and two focus group discussions, five attributes and their appropriate levels were identified to characterise heritage site preservation policies: cultural heritage site category, levy time horizon, annual visit rate (as a percentage) to the heritage sites, heritage distance from residences dwelling and annual levy specific to heritage site category. The payment was presented as a household levy that would apply to all Shiraz households. Participants were informed that the proposed levy would be applied to property or passed along in the form of increased rental payment (the latter made explicit with the intent of informing renters that they would indirectly be affected). The levy would be imposed for a specific time duration, ranging from 10-50 years. The literature indicates that one-off payments can be excessively conservative, which led to our use of the annual levy (MacDonald, Ardeshiri, Rose, Russell, & Connell, 2015; Whitehead & Blomquist, 2006). To arrive at a reasonable range of levies to test, we used an estimate of the net present value of total current households in Shiraz (plus 25% for upper bond and -25% for a lower bound) and a 3% annual interest rate over 50 years, say, to calculate upper and lower levy amounts for heritage sites. Figure 5 provides an overall upper and lower range of levies, where a sufficiently wide range allows coverage for a comprehensive set of future analyses, in terms of population affected by the levy. Table 1 presents the full list of attributes and levels considered for the referendum task.

**Figure 5:** Wide range of levies used for estimating willingness to pay for heritage presevaition

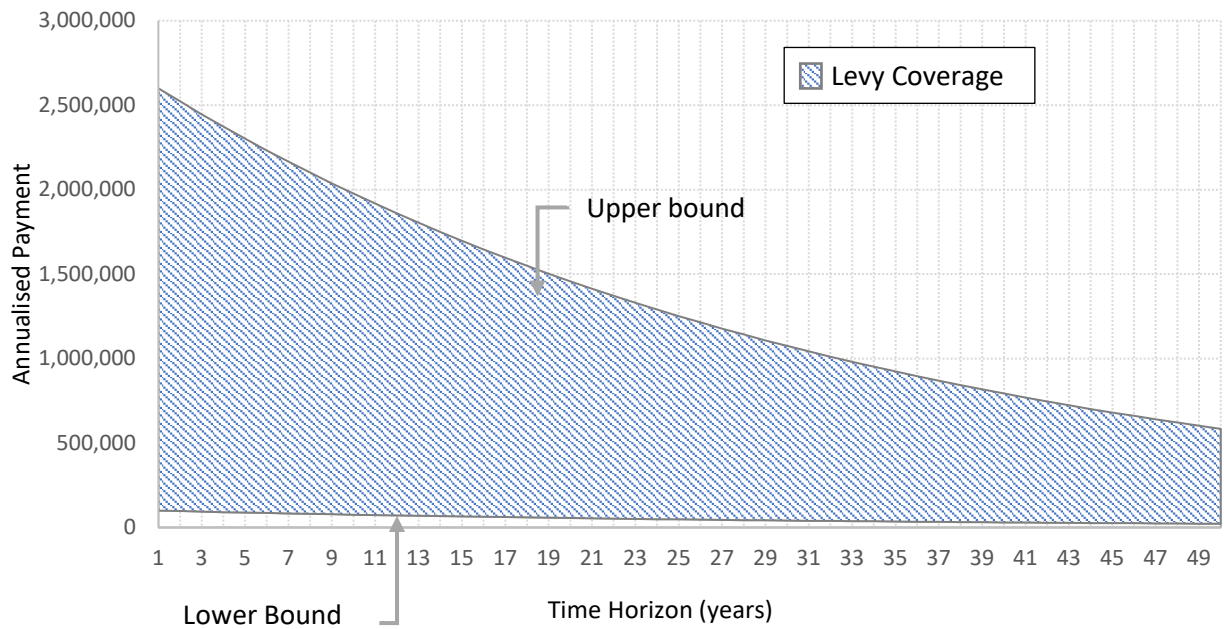

**Table 1:** Attributes and levels used in the referendum task

| Attributes | levels |
|---|---|
| Cultural heritage category | Historical sites, Religious Sites, Gardens |
| Time horizon | 10 years, 20 years, 40 years, 50 years |
| Decrease in annual visit rate (%) | 5%, 10%, 15%, 20% |
| Distance from residential location | 1km, 10km, 25km, 50km |
| Annual levy to your household (*in Rials, n*ote that at the time of data collection 1USD = 32,415 Rials) | 100,000, 250,000, 500,000, 750,000, 1,000,000, 1,500,000, 2,000,000, 2,500,000 |

## 3.3 EXPERIMENTAL DESIGN

Individual policy preferences were measured using the choice modelling framework presented in Figure 6. Individuals could select between status quo (with no specific preservation action taken, leading to a deterioration in the condition of the site and a decrease in annual visitor numbers) and a proposed levy to pay for management that prevents visitation loss and maintains the current condition of the heritage site. As mentioned earlier, three heritage categories (*Historical*, *Religious*, *Gardens*) were studied. Respondents were asked to choose between two options (see Figure 7):

**Yes**: for a given heritage site of a certain nominated category and specific proximity to their residential location, an annual levy of the amount shown and for the time horizon specified would be used to preserve the site and maintain its current visitation rate.

**No**: this 'status quo' alternative meant that the heritage site received no specific incremental preservation action taken by the local council. The consequence of voting for this policy was that the specific heritage would suffer a loss of visitation over the time horizon ("no specific maintenance action taken" policy scenario on the left in Figure 7). Residents of Shiraz would not pay any extra levy in this case.

**Figure 6:** Referendum task structure

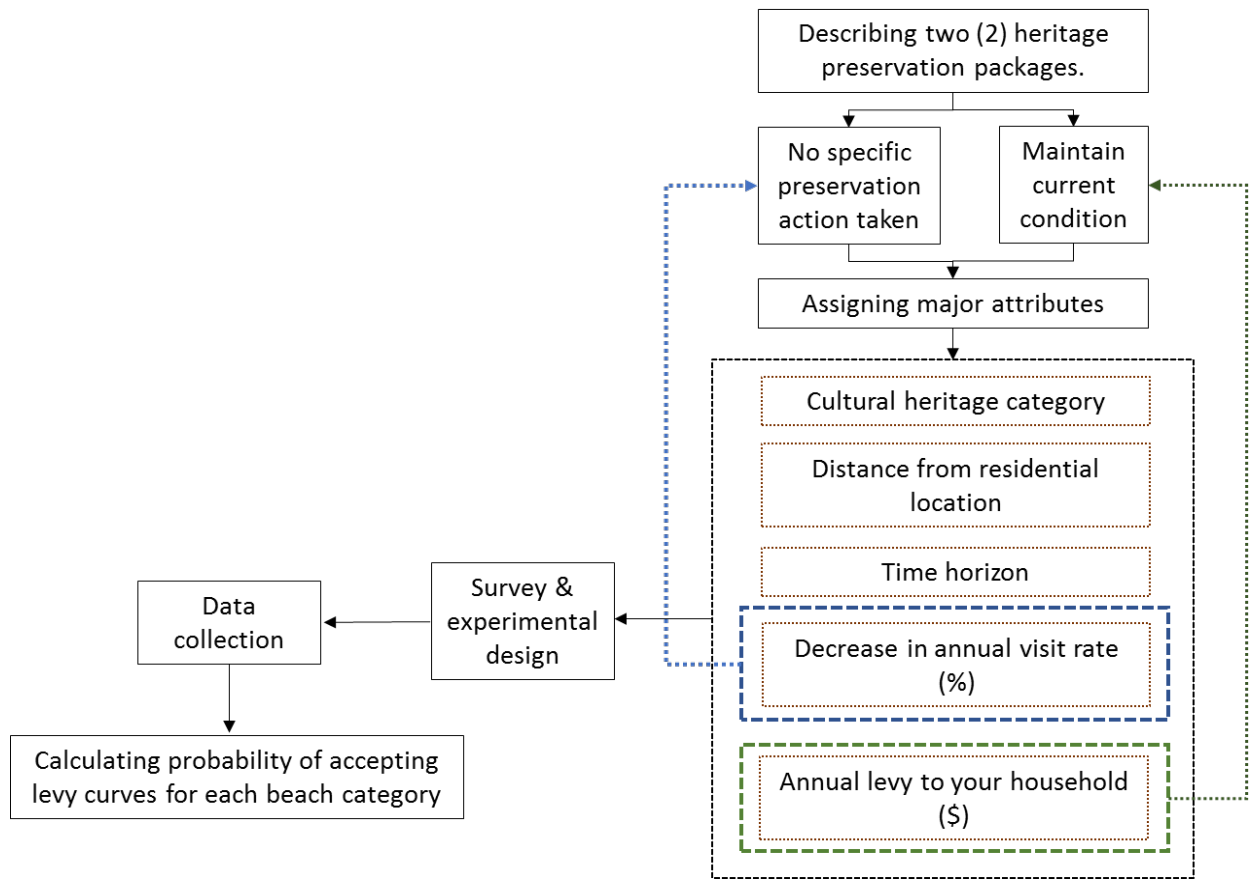

The attribute level values used in specific choice tasks were defined by an orthogonal main effects experimental design (J. Louviere, D. A. Hensher, & J. D. Swait, 2000). Ngene software was used to generate the design for this study. The final design included 48 choice tasks in 8 blocks providing each participant with 6 repeated choice occasions. In completing each hypothetical referenda, the individual was urged to treat each referenda independently of the others. Participants were also reminded to keep in mind their available household income and all other things that this income is spent on. To ensure that the participants took the survey seriously, a short cheap talk script was developed using guidance from Morrison and Brown (2009). Cheap talk is a technique used in SP surveys to remind participants that they should make choices as if they really had to pay. Cheap talk has been shown to be effective at reducing the potential for hypothetical bias in choice experiments (List, Sinha, & Taylor, 2006; MacDonald, Ardeshiri, Rose, Russell, & Connell, 2015; Tonsor & Shupp, 2011). Figure 7 presents an example of the referendum task.

**Figure 7**: An example of the referendum task

|  | No specific preservation action taken | To preserve and restore current condition |
|---|---|---|
| Cultural heritage site | *A Historical site* | |
| Time horizon | 10 years | |
| Annual visit rate | 10 % decrease in annual visit rate | Maintain its current visit rate |
| Distance from your home | 10 Km | |
| Annual levy | 0 | 50,000 Toman for 10 years |
| I would vote for this option: *Choose one only* | ☐ | ☐ |

### 3.4 DATA COLLECTION

Data for our analysis came from a state-wide sample of Shiraz residents. The survey was administered in March 2017 through face to face interviews. The sample is reasonably representative of Shiraz city residents. The survey was paper-based and conducted using trained interviewers. Respondents were randomly sampled based on the relatively equal distribution of the population within Shiraz. The survey was directed at decision makers in the household, i.e. those more likely to pay the rent and property expenses. Overall 489 respondents were interviewed for this study.

Sample characteristics are given in Table 2. From the total 489 sampled population, there is a higher male participation (59.3%) compared to Female (40.7%). The majority of participants were in the 18-34 age bracket (45%), 34% were in the 35-64 age bracket and the remaining 21% were aged a65 years and above. The average age was 37.3 years. Participants were from different types of households with the majority (37.8%) being "Couple family with children" with the average household size of 3.11 per household.

The common education level among the sampled population was "bachelor's degree" (45.3%). The average household income was just above 34 million Rials (equivalent to USD$1057) with majority

following in the 25 to 45 million income bracket. Homeowners constitute 68.9% of the sample, and renters the remainder with majority living in a free-standing house (66.5%). Respondents were mostly from the metro area (78%) and the remaining 22% were from the regional.

Of the 489 respondents, 87 (17.7%) had mentioned that they have not visited a heritage site in the past 12 months, nor are willing to visit a one: in this study, such respondents are classified as not being heritage visitors (non-user). The remaining 402 (82.3%) indicated that they have visited a heritage site in the past 12 months and are considered as heritage visitors (user). This latter group reported that in aggregate they have made 4477 visits to a pre-specified set of 3 nominated heritage sites, resulting in an average of 9 visits per year per household. From the 4477 visits, 1783 were to historical sites, 1393 to gardens and 1301 to religious sites.

**Table 2:** Descriptive statistics of respondents

| Variable | | Statistics |
|---|---|---|
| *Total Participants* | | *489* |
| *Gender* | | |
| | Male | 59.3% |
| | Female | 40.7% |
| *Age* | | |
| | Sampled average | 37.3 years |
| | Age bracket (18 -34 years) | 45% |
| | Age bracket (35 -64 years) | 34% |
| | Age bracket (65 years and above) | 21% |
| *Household type* | | |
| | Couple family with no children | 27.2% |
| | Couple family with children | 37.8% |
| | One parent family | 15.5% |
| | Single person household | 10.5% |
| | Group household | 9% |
| | Other Family | |
| *Household size* | | |
| | Average household size | 3.11 |
| *Education* | | |
| | College graduate or less | 23.7% |
| | Bachelor's degree | 45.6% |
| | Master's or PhD degree | 30.7% |
| *Household annual income* | | |
| | Average Income | 34,279,000 Rial (USD $1057.5) |
| | Income bracket (below 25 million Rial) | 36% |
| | Income bracket (25 to 45 million Rial) | 40.7% |
| | Income bracket (above 45 million Rial) | 23.3% |
| *Dwelling type* | | |
| | Free standing house | 66.5% |
| | Semi-detached, in a row of terrace houses, townhouse | 10.8% |
| | Flat, unit or apartment | 22.0% |
| | Other dwelling (e.g. caravan, cabin, houseboat, or improvised home) | 0.7% |
| *Is this dwelling…?* | | |
| | Owned | 68.9% |
| | Rented | 31.1% |
| *Region* | | |
| | Metro | 78.1% |
| | Regional | 21.9% |
| *Visited a heritage site in the past 12 months* | | |
| | Yes | 82.3% |
| | No | 17.7% |
| *Heritage visit* | | |
| | Total visit per year | 4477 visits |
| | Average visit per year per household | 9.1 visits |
| | Visited a historical site | 40% |
| | Visited a garden | 31% |
| | Visited a religious site | 29% |

# 4. DATA ANALYSIS

A concern with hypothetical referendum tasks is the possibility of strategic or protest voting in the form of "Nay-saying" and "Yea-saying" (i.e., voting 'no' irrespective of policy attributes variation, and voting 'yes' no matter the policy attributes). Among psychologists and sociologists studying response acquiescence, yea-saying is defined as the tendency to agree with questions regardless of content. Vice versa the tendency to disagree is defined as Nay-saying (Blamey, Bennett, & Morrison, 1999; Loomis, 2014; Moum, 1988). Traditional statistical analyses of DCEs do not handle these extreme preferences well. Recognising this limitation, the random utility choice model utilised for this study is based on an innovative use of a standard Latent Class (LC) model.

To begin, the proposed model allows the sample to be separated between those who make trade-offs and those who don't; among those who don't make trade-offs, it makes a distinction between those unswervingly protesting against or in favour of the referenda. For those who make trade-offs, it is assumed that the individuals may be decomposed into discrete segments that differ in their predisposition towards heritage maintenance policy and their sensitivity to different attributes presenting the policy. Thus, in addition to handling trade-off heterogeneity among "traders", we allow one segment to represent the Yea-sayers and another segment to represent the nay-sayers.

Figure 8 illustrates a path diagram of the underlying structural model representing the choice process. Sociodemographic characteristics and individual's choice behaviour in response to a given policy set, are the observable – or manifest - variables (presented in rectangular shapes). Following Swait (1994) we allow sociodemographic characteristics as well as individual perceptions, knowledge and experience of a given heritage site to form the "segment membership" as well as "taste in preference" in residents' choices. Structural latent variables are depicted through ellipses.

(1) S*ociodemographic characteristics* form the latent segment membership likelihood functions for an individual.

(2) Through a latent *segment classification mechanism,* the membership likelihood functions determine the latent segment (i.e. yea-sayers, nay-sayers and traders) to which an individual belongs.

(3) The decision-maker has preferences with respect to the policy which determine the yes/no vote. These preferences are determined by the individual's perceptions, knowledge and experience of a given heritage site, sensitivity to the given attributes, his/her personal characteristics and the latent class to which he/she belongs. These preferences are conditional on, and specific to, the segment to which the person belongs.

This structural model is an adaptation of the general framework presented in McFadden (1986) and Swait (1994). The latent class model (LCM) has been used extensively for the analysis of individual heterogeneity (for theoretical discussion see Boxall & Adamowicz, 2002; Greene & Hensher, 2003; Swait & Adamowicz, 2001).

**Figure 8:** A structural equation model of latent segmentation and choice process. Partially adapted from Figure 1, Swait (1994).

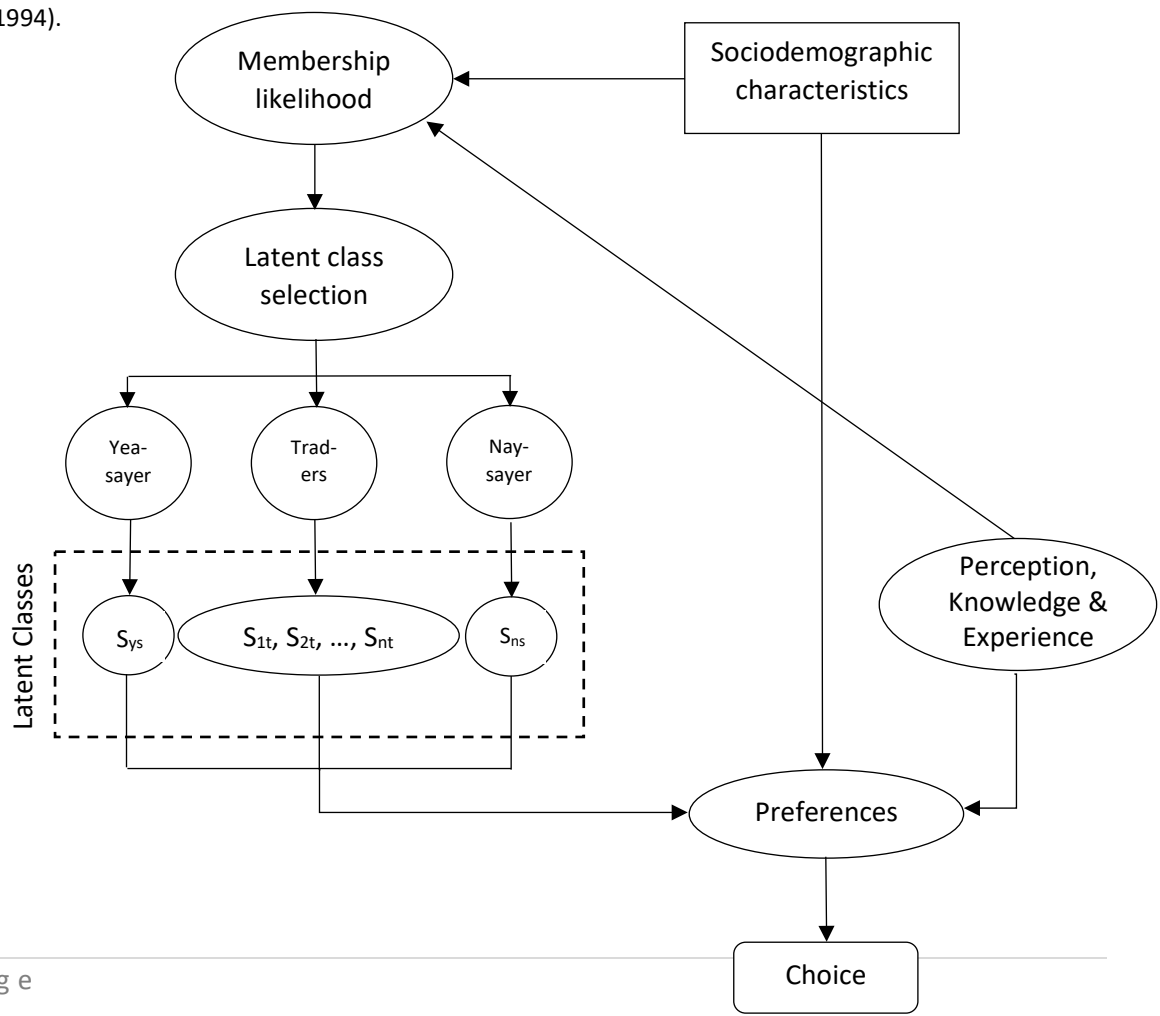

## 4.1 MODEL FORMULATION

The underlying theory of the LCM posits that individual behaviour depends on observable attributes and on latent heterogeneity that varies with factors that are unobserved by the analyst. LCM will probabilistically segment the sample to be homogeneous within and heterogeneous across segments, with respect to the choice process. We propose to analyse the heterogeneity through a model of discrete parameter variation. Thus, it is assumed that individuals are implicitly sorted into a set of *S* classes (whether known or not to that individual), but which class contains any particular individual is unknown to the analyst. Having said that, by definition, we expect from the yea-sayers - who always agree to pay a levy to maintain heritage visit rate regardless of the costs and benefits – to be deterministic and have probability value equal to one for the Yes alternative. For this reason, we allow the utility for paying a levy ($U_{ivs}$) to be a very large positive value (effectively, positive infinity) for this segment. Vice versa, by fixing $U_{ivs}$ to a large negative value (effectively, negative infinity) we force the response of individuals in this segment to be No with probability one; conversely, the nay-saying segment has zero probability of voting Yes. The segment(s) who make trade-offs between the given options are considered to have a finite (and to be estimated) $U_{ivs}$ as specified in equation (1).

$$U_{ivs} = \begin{cases} \text{If } \textit{Yea-sayers} & U^*_{i|s} = +\infty \\ \text{If } \textit{Trader} & -\infty < U^*_{i|s} < +\infty \\ \text{If } \textit{Nay-sayers} & U^*_{i|s} = -\infty \end{cases} \qquad (1)$$

To establish certain components required to build the model, we assume that a choice scenario presents alternatives in choice set C_r, r=1,…,R, where R is the number of choice scenarios in a choice experiment. Each alternative *i* has utility

$$U_{ir|s} = V_{ir|s} + \varepsilon_{ir|s}, i \in C_r, \qquad (2)$$

where $V_{ir|s}$ is the systematic utility for the alternative in the r[th] scenario, conditional on belonging to class $s$ (=1,…,S) with a set of preference component $\beta_s$ and an individual's sociodemographic characteristics $Z_i$ as well their perception, experience and knowledge $\gamma_i$ as such that

$$V_{ir|s} = \beta_s X_{ir|s} + Z_i \sigma_{ir|s} + \gamma_i \rho_{ir|s}, \qquad (3)$$

and $\varepsilon_{ir|s}$ is the stochastic utility of the alternative. If we assume that the $\varepsilon_{ir|s}$ is independently and identically Gumbel distributed with scale $\mu$, the class conditional choice model is a multinomial logit formulation:

$$P_{ir|s} = \frac{\exp(\mu V_{ir|s})}{\sum_{j \in C_r} \exp(\mu V_{jr|s})} . \qquad (4)$$

We assume that given the class assignment, the $R_i$ events are independent. This is possibly a strong assumption, especially given the nature of the sampling design used in our application—a stated choice experiment in which the individual answers in sequence, and in short order, repeated choice scenarios. In fact, there might well be correlation in the unobserved parts of the random utilities. The latent class does not readily extend to deal with this potential autocorrelation, so we have left this aspect for further research. Thus, for the given class assignment, the contribution of individual *i* to the likelihood would be the joint probability of the sequence $P_i$=[ $P_{i1}$, $P_{i2}$,…, $P_{ir}$]. This is

$$P_{i|s} = \prod_{r=1}^{R} P_{ir|s} \qquad (5)$$

The class assignment is unknown. Let $H_{is}$ denote the prior probability for class *s* for individual *i*. The polytomous multinomial logit form is

$$H_{is} = \exp(z_j' \theta_s) / \sum_{s=1}^{S} \exp(z_j' \theta_s) , s = 1, \ldots, S, \theta_s = 0, \qquad (6)$$

where $z_i$ denotes a set of observable characteristics which enter the model for class membership. Note that not all θ's can be identified since the corresponding variables do not vary from class to class.

Hence, one must normalize one of these vectors to a constant, say, zero. We have chosen to normalize the θ for the last class, S. The likelihood for individual *i* is the expectation (over classes) of the class-specific contributions:

$$P_i = \sum_{s=1}^{S} H_{is} P_{i|s} \tag{7}$$

The Log likelihood for the sample is

$$lnL = \sum_{i=1}^{N} \ln P_i = \sum_{i=1}^{N} ln \left(\sum_{s=1}^{S} H_{is} \prod_{r=1}^{R} P_{ir|s}\right) \tag{8}$$

Maximization of the log likelihood with respect to the *S* structural parameter vectors, $β_s$ and the *S*–1 latent class parameter vectors, $θ_s$ is a conventional problem in maximum likelihood estimation.

## 5. RESULTS

### 5.1 UNDERSTANDING PREFERENCES OF DIFFERENT SURVEY SEGMENTS

A latent class model was used to estimate individual policy preferences for heritage site preservation to maintain its current visitation rate. In determining the final model, numerous models were estimated where we tested different utility specification and number of classes. Coincidentally, based on a comparison across both statistical measures of fit, such as the Bayesian Information Criterion (BIC) and the Akaike Information Criterion (AIC), and behavioural interpretation, a five-class model was selected as the preferred model specification. To ease comprehension, we've named each segment an ordered them in terms of how strategically they have respondent to the referendum task.

Segment one was *"Yea-sayers"*, who said *yes* to any amount of levy payment for any heritage sites and the second segment was *"Nay-sayers"*, who were unwilling to pay any levy at all, no matter the amount or heritage type. The third segment was the *"Historical site yea-sayers"* who were willing to pay any levy amount for historical sites only. These three segments have completely responded to the referendum task strategically. This behaviour has been widely acknowledged in environmental choice modelling literature (see for example Bateman, Langford, Turner, Willis, & Garrod, 1995; Bennett &

Adamowicz, 2001; Bennett & Blamey, 2001). Segment four was the "*Religious site nay-sayers*" who were partially strategic with their responds. This segment was unwilling to pay any levy amount for religious sites, however, made trade-off for any historical and gardens site. Finally, the fifth group was the *"Traders"* who basically were completely engaged with trading-off between the given options.

To understand the relative distribution of the five respondent segments within the sampled population, a sample enumeration was carried out to calculate the expected size of each class. As presented in Table 3, in total 44 % of the sample were categorised as being voting *completely strategic* in favour of or protesting the given policy. Among those, the model predicts that 22.7% of the population belong to the yea-saying, 13.9% belong to the nay-saying and 7.4% to the historical site yea-saying segment. More interestingly, 36.8% were *partially strategic* with their voting and only 19.2% of the sample voted *non-strategically*.

**Table 3.** Population segments and associated membership probabilities

| | Completely strategic | | | Partially strategic | Non-strategic |
|---|---|---|---|---|---|
| ***Segments*** | Yea-sayers | Nay-sayers | Historical site yea-sayers | Religious site nay-sayers | Traders |
| ***Segment rule*** | Always accepts to pay a levy | Never accepts to pay a levy | Always accepts to pay a levy for **historical sites** only. | Never accepts to pay a levy for **religious sites** only. | Makes trade-offs between the given scenarios. |
| ***Segments propensity*** | 22.7% | 13.9% | 7.4% | 36.8% | 19.2% |

Table 4 provides the parameter estimation results for each of the segments. These segments differ from each other in terms of their sensitivity to different policy attributes, and their demographic characteristics. Over subsequent paragraphs, we summarize some of the key attributes of each of these five classes to underscore the behavioural differences between the segments. Linear and non-linear (quadratic) transformation of the continuous variables were both expressed in the utility functions. Further, to avoid collinearity of the linear and quadratic terms, *orthogonal polynomial* coding was used (for more information refer to chapter 9 in Louviere, Hensher, & Swait, 2000).

**Segment 1 - Yea-sayers:** This segment is unlikely to be in the lower age spectrum (aged between 18-34 years). They are a family couple with no children and associated with a higher index score in the 'experience index[1]'. This index includes an average scoring of responding to a 5-point Likert scale related to how the individual felt regarding their last visit to a heritage site. The scores ranged from strongly disagree (score of 1) to strongly agree (5) for the following set of statements; had fun; were entertained; spent quality time with family and friends and found the heritage site pleasant. This segment has indicated that they are willing to pay any amount of levy within the studied range (100,000 -2,500,000 Rial) to maintain the current visitation rate for any heritage sites for any given time horizon and distance to heritage site.

**Segment 2 - Nay-sayers**: It is expected that this class is largely constituted of relatively young female in a one parent family structure with a college or below education level. They are unlikely to reside in the metro area and their total household income is associated with the lowest range of household income spectrum. Increase in the household size also increases the possibility of belonging to this segment. As mentioned previously this segment are unwilling to pay any levy at all, no matter the amount or heritage type.

**Segment 3 - Historical site yea-sayers:** The characteristics associated with the class membership of this segment are those of affluent households. They are more likely to reside in the metro area and have a master's or a PhD degree, with the total household income of more than 45 million Rial. Households belonging to this segment are more likely to be a couple with children, and be in the age bracket of 35-64 years old. As part of the survey we enquired about respondents' awareness sets regarding each heritage category studied. For this reason, each household was presented with each of the categories (i.e. garden, religious and historical) and were asked to indicate if they have visited the site previously. Households belonging to this segment are associated with indicating that they had visited the majority of the historical sites previously. This segment indicated that they are willing to pay any amount of levy within the studied range (100,000 -2,500,000 Rial) to maintain the current visitation rate for the historical sites only, and they are unwilling to pay any levy amount for gardens and religious sites no matter what the policy offered.

---

[1] This index is defined based on "non-mindful benefits" thought process described in (McIntosh, 1999)

**Segment 4 - Religious site nay-sayers:** Similar to segment (3), households belonging to this segment are more likely to reside in the metro area and have a master's or a PhD degree. They are likely to be a one parent household in the age bracket of 35-64 years old. They are unlikely to have a high household size and an education level of college level or less.

This segment are unwilling to pay any levy amount for the religious sites no matter what the policy offers, however, they make trade-offs between policies related to historical and garden sites. In terms of their preferences towards the attributes of the latter sites, they have similar preferences towards time horizon and reduction in annual visit rates for both the sites. However, they are relatively more sensitive toward these attributes for the gardens compared to the historical sites. For a higher time, horizon and/or a lower percentage in reduction to the annual total visit, their probability of willing to pay levy decreases. This segment is indifference towards their distance from the gardens, however, their probability for WTP levy reduces if distance from the historical sites goes towards both end of distance spectrum (i.e. if the respondent is too close or too far away). Meaning that the WTP curve based on distance from site (holding every other variable constant) has a concave shape with its maximum value in the mid-point. Finally, with an increase in the levy amount, the probability of accepting the policy decreases. Having said that, the household belonging to this segment are more sensitive towards an increase in the levy amount for gardens compared to the historical sites.

**Segment 5 – Traders:** This segment was the base group for estimating the membership propensity function. Based on their taste in the preference function, belonging to the 45-64 years age bracket, frequency of visit to religious site and being a couple family with no children increases the probability of willing to pay a levy, whereas, being a one parent family reduces this probability. People in this segment are indifferent to the time horizon and reduction of visit rates for historical sites. Similar to households belonging to segment 4, their probability for WTP levy reduces if distance from the historical sites goes towards both end of the distance spectrum (i.e. 1km and 50km); and also with an increase in the levy amount, the probability of accepting the policy decreases. Interestingly this segment, when it comes to religious sites, has a higher preference towards both ends of the time horizon spectrum (WTP curve has a convex shape), meaning that they are more likely to accept paying a levy for a very short or very long duration. Even more surprising, the WTP curve based on reduction in annual visit rate has a concave shape and they are less likely to accept the policy if the reduction in the visit rate is too little or too high. An explanation for being unwilling to pay for the high reduction in the visit rate is that majority of the religious heritage sites are still functioning and provide services for any visitor to practice their religious observances on site. This may provide a circumstance that the household can benefit to practice his/her religious more freely in a less crowded environment. Finally, with an increase in proximity to a religious site WTP probability decreases. The same also applies for increase in the levy amount. This segment has the lowest sensitivity towards any changes regarding the latter two attributes discussed for religious sites relative to the historical and garden sites. Regarding their taste preferences for the garden sites, this segment is very sensitive toward an increase in time horizon (relative to the other two heritage categories). The WTP decreases for an increase in time horizon. The preference for reduction in annual visit rate is similar to the religious site and has a quadratic function. Based on the sign of estimated parameter, the WTP curve for reduction in annual visit rate has a concave shape and the probability of accepting the policy for a very low or high reduction in visitation rate, decreases. Privacy related issues maybe an explanation for the dis-utility of accepting to pay a levy for the high reduction rate as the nature of the garden site and the flowers can provide a romantic environment. The distance attribute in the utility function have a quadratic function, and therefore the probability for WTP curve based on distance from garden site has a concave shape with its maximum value in the mid-point. Finally, this segment is more sensitive towards paying any levy amount for religious sites in comparison to the other two sites (historic and garden) and their WTP decreases by an increase in the levy amount (please refer to table for all the estimated parameters magnetite, sign and significance).

**Table 4:** Parameter estimates for the referendum task

| | Parameters | Value | Std err | t-test | p-value |
|---|---|---|---|---|---|
| **Yea-sayers** | | | | | |
| | Constant | | Fixed (at +∞) | | |
| **Nay-sayers** | | | | | |
| | Constant | | Fixed (at -∞) | | |
| **Traders** | | | | | |
| | Constant | -1.35 | 0.308 | -4.39 | <0.001 |
| *Historical sites* | | | | | |
| | Constant | 2.09 | 0.624 | 3.35 | <0.001 |
| | Time horizon | 0 | 0 | 0 | |
| | Annual visit rate reduction | 0 | 0 | 0 | |
| | Distance to the site (quadratic form) | -1.4 | 0.458 | -3.06 | <0.001 |
| | Levy | -0.788 | 0.206 | -3.82 | <0.001 |
| *Religious sites* | | | | | |
| | Constant | | Fixed (at zero) | | |
| | Time horizon (quadratic form) | 1.09 | 0.329 | 3.31 | <0.001 |
| | Annual visit rate reduction (quadratic form) | -0.685 | 0.225 | -3.05 | <0.001 |
| | Distance to the site | -0.153 | 0.0983 | -1.56 | 0.12 |
| | Levy | -0.281 | 0.0701 | -4 | <0.001 |
| *Gardens* | | | | | |
| | Constant | -2.04 | 0.842 | -2.43 | 0.02 |
| | Time horizon | -4.68 | 1.5 | -3.11 | <0.001 |
| | Annual visit rate reduction (quadratic form) | -4.2 | 1.4 | -3.01 | <0.001 |
| | Distance to the site (quadratic form) | -2.97 | 1.17 | -2.53 | 0.01 |
| | Levy | -1.74 | 0.6 | -2.91 | <0.001 |
| *Demographics* | | | | | |
| | Age bracket (35 -64 years) | 0.747 | 0.44 | 1.7 | 0.09 |
| | Frequency of visit to Religious sites | 0.088 | 0.0445 | 1.97 | 0.05 |
| | Couple family with no children | 3.04 | 1.05 | 2.89 | 0 |
| | One parent family | -1.8 | 0.872 | -2.07 | 0.04 |
| **Religious site nay-sayers** | | | | | |
| | Constant | 6.92 | 0.443 | 15.62 | <0.001 |
| *Historical sites* | | | | | |
| | Constant | -4.41 | 0.486 | -9.07 | <0.001 |
| | Time horizon | -0.318 | 0.0966 | -3.29 | <0.001 |
| | Annual visit rate reduction | 0.986 | 0.145 | 6.79 | <0.001 |
| | Distance to the site (quadratic form) | -0.323 | 0.112 | -2.9 | <0.001 |
| | Levy | -0.259 | 0.0395 | -6.56 | <0.001 |
| *Religious sites* | | | | | |
| | Constant | | Fixed (at zero) | | |
| | Time horizon | | | | |
| | Annual visit rate reduction | | *Fixed (at -∞)* | | |
| | Distance to the site (quadratic form) | | | | |
| | Levy | | | | |
| *Gardens* | | | | | |
| | Constant | -7.36 | 0.495 | -14.88 | <0.001 |
| | Time horizon | -0.449 | 0.147 | -3.05 | <0.001 |
| | Annual visit rate reduction | 1.24 | 0.213 | 5.8 | <0.001 |
| | Distance to the site (quadratic form) | 0 | 0 | 0 | <0.001 |
| | Levy | -0.705 | 0.0908 | -7.76 | <0.001 |
| *Demographics* | | | | | |
| | Awareness of number of heritage sites available | 0.0894 | 0.0494 | 1.81 | 0.07 |
| | Couple family with children | -0.874 | 0.358 | -2.44 | 0.01 |
| | Couple family with no children | -1.09 | 0.448 | -2.43 | 0.02 |
| **Historical site yea-sayers** | | | | | |
| *Historical sites* | | | | | |
| | Constant | | Fixed (at +∞) | | |
| *Religious sites* | | | | | |
| | Constant | | Fixed (at -∞) | | |
| *Gardens* | | | | | |
| | Constant | | Fixed (at -∞) | | |

**Table 4:** Continue

| | Segment membership | | | | |
|---|---|---|---|---|---|
| | Parameters | Value | Std err | t-test | p-value |
| **Yea-sayers** | | | | | |
| | Constant | -2.24 | 0.526 | -4.25 | 0 |
| | Age bracket (18 -34 years) | -1.42 | 0.433 | -3.28 | 0 |
| | Couple family with no children | 4.28 | 0.425 | 10.07 | 0 |
| | Experience index | 0.304 | 0.116 | 2.63 | 0.01 |
| **Nay-sayers** | | | | | |
| | Constant | -9.12 | 2 | -4.56 | 0 |
| | Age bracket (18 -34 years) | 2.85 | 1.01 | 2.82 | 0 |
| | One parent family | 3.68 | 1.41 | 2.61 | 0.01 |
| | College graduate or less | 2.74 | 0.859 | 3.18 | 0 |
| | Income bracket (below 25 million Rial) | 1.47 | 0.758 | 1.94 | 0.05 |
| | Living in metro | -2.48 | 0.781 | -3.17 | 0 |
| | Female | 1.46 | 0.798 | 1.83 | 0.07 |
| | household size | 0.912 | 0.297 | 3.07 | 0 |
| **Traders** | | | | | |
| | Constant | Based group (fixed at zero) | | | |
| **Religious site nay-sayers** | | | | | |
| | Constant | 1.11 | 0.854 | 1.3 | 0.19 |
| | Age bracket (35 -64 years) | -1.14 | 0.474 | -2.42 | 0.02 |
| | One parent family | -2.46 | 0.788 | -3.13 | 0 |
| | College graduate or less | -1.22 | 0.482 | -2.53 | 0.01 |
| | Master's or PhD degree | 0.53 | 0.374 | 1.42 | 0.16 |
| | Living in metro | 1.38 | 0.724 | 1.91 | 0.06 |
| | Household size | -0.477 | 0.131 | -3.64 | 0 |
| **Historical site yea-sayers** | | | | | |
| | Constant | -45.1 | 19.5 | -2.31 | 0.02 |
| | Age bracket (35 -64 years) | 4.98 | 3.22 | 1.55 | 0.12 |
| | Couple family with children | 2.29 | 1.77 | 1.29 | 0.2 |
| | Master's or PhD degree | 3.44 | 2.09 | 1.65 | 0.1 |
| | Income bracket (above 45 million Rial) | 4.37 | 2 | 2.18 | 0.03 |
| | Living in metro | 3.52 | 1.95 | 1.81 | 0.07 |
| | Historical site awareness | 4.11 | 2.05 | 2 | 0.05 |
| **Estimation report** | | | | | |
| | Number of estimated parameters | **56** | | | |
| | Sample size | **2934** | | | |
| | Log likelihood | **-607.58** | | | |

## 5.2 WILLINGNESS TO PAY FOR DIFFERENT HERITAGE SITES

Willingness to pay for management that preserves the heritage site and prevents visitation loss differed by segment and by heritage site (Table 5). A sample enumeration was carried out to calculate the WTP amounts. The household average WTP amount was calculated using the sum of all five segments population propensity multiplied by the average WTP levy amount for each segment. Overall the WTP was highest for 'Historical sites' (1,170,030 Rials). WTP for garden and religious site were calculated respectively at 883,808 Rials and 710,934 Rials. These WTP figures represent the sample population average values, whereby the proportion of households in the city that would be considered as a yea-sayers, nay-sayers, historical site yea-sayers, religious site nay-sayers and traders (based on their demographic characteristics) needs to be weighted up to the entire population of the city to be able to represent the true average WTP per household.

**Table 5:** Average household willingness to pay for management that preserve heritage sites, by heritage site and population segment

|  | Household average WTP (Rial per annum) | Yea-sayers* (Rial per annum) | Nay-sayers (Rial per annum) | Historical site yea-sayers (Rial per annum) | Religious site nay-sayers (Rial per annum) | Traders (Rial per annum) |
|---|---|---|---|---|---|---|
| ***Historical sites*** | 1,170,030 | 2,500,000 | - | 904,170 | 940,170 | 987,700 |
| ***Religious sites*** | 710,934 | 2,500,000 | - | - | - | 747,050 |
| ***Gardens*** | 883,808 | 2,500,000 | - | - | 572,090 | 550,930 |

\* For yea-sayers, WTP is dependent on the upper bound of levy amount (see Table 1)

# 6. DISCUSSION

Results of the referendum choice experiment presented in this study provide an estimate of WTP for management to preserve heritage sites in the face of erosion and to maintain current visitation rate. We find that 86% of the population would be willing to pay some levy amount, dependent on the policy setting. Like other CV and CM studies, our study provides an estimate of non-market value. Given that our referendum questioned respondents' willingness to pay a levy irrespective of heritage visitation, we consider that it represents non-use value. Moreover, because it was clear in our survey that any levy would be additional to the travel costs that individual respondents would incur to access a given heritage site, we consider that the non-use values estimated in this study are additional to use (e.g. recreation) values. The non-use values associated with the preservation of heritage sites estimated in our study can be used as an input to cost-benefit analysis of heritage management options in Shiraz (and elsewhere if benefit transfer is undertaken in an appropriate manner). This would enable the economic outcomes associated with different configurations of built and natural assets to be assessed and optimised in line with local community preferences for future heritage site configuration.

Our results indicate that there is a strong 'preservationist' attitude, whereby survey respondents demonstrated a preference to maintain the condition of heritage sites and visitation rates in their current level. This can be considered a positive result in terms of a sustainable financing perspective, in that WTP spans the breadth of the heritage sites and is not merely a reaction to intense heritage risk or damage, but it presents difficulties in terms of strategic prioritisation.

However, other findings can assist with spatial prioritisation of heritage management. These include the finding that respondents were willing to pay different levy amounts for different heritage sites (historical, religious and gardens). We also find differences in WTP amongst different populations segments (yea-sayers, nay-sayers, historical site yea-sayers, religious site nay-sayers and traders). To the extent that these can be linked to socio-demographic characteristics as well as awareness and experience, they can also be used to discriminate the value and heritage preferences of a specific community or local government area in order to assist with spatial prioritisation.

As a final note on the nature of non-market values for heritage sites presented in this study, we highlight that it is not necessarily the case that *nay-sayers* hold a zero non-use value for heritage sites. An alternative explanation is that they may have lodged a 'protest vote' about the proposed payment vehicle (annual levy) or about where responsibility for further investment in heritage preservation lies – they may think that it should already be covered in their taxes. The size of this proportion of the population (14%) may be of concern if decision-makers are seeking to implement a mandatory levy to support heritage management. We recommend further research to identify respondents' motivations for nay-saying in order to determine if a more acceptable payment vehicle might be conceived.

## 7. CONCLUSIONS

Many authors have reported that the funds for environmental preservation and management of natural or protected areas are insufficient and declining (Banhalmi-Zakar, Ware, Edwards, Kelly, & Becken, 2016; Baral, Stern, & Bhattarai, 2008; Dharmaratne, Sang, & Walling, 2000; Eagles, McCool, Haynes, Phillips, & Programme, 2002; Lindberg, 1998; Reynisdottir, Song, & Agrusa, 2008). Our study suggests there is a WTP within the community to preserve heritage sites in the face of future erosion and to maintain current visitation rates, raising the possibility of realising better funding arrangements for the preservation of heritage assets.

Our findings contribute to the current literature by providing significant empirical findings that heritage managers can use in their decision making as well as investigating a new public funding mechanism. Moreover, from a methodological perspective, this study is innovative in using a latent class model in the treatment of strategic or protest voting in the form of "nay-saying" as well as "yea-saying" at the estimation stage rather than through elimination by the researcher prior to the estimation.

# REFERENCE

Aas, C., Ladkin, A., & Fletcher, J. (2005). Stakeholder collaboration and heritage management. *Annals of Tourism Research, 32*(1), 28-48.

Ardeshiri, A., Willis, K., & Ardeshiri, M. (2018). Exploring preference homogeneity and heterogeneity for proximity to urban public services. *Cities*. doi:https://doi.org/10.1016/j.cities.2018.04.008

Banhalmi-Zakar, Z., Ware, D., Edwards, I., Kelly, K., & Becken, S. (2016). Mechanisms to finance climate change adaptation. In: National Climate Change Adaptation Research Facility.

Baral, N., Stern, M. J., & Bhattarai, R. (2008). Contingent valuation of ecotourism in Annapurna conservation area, Nepal: Implications for sustainable park finance and local development. *Ecological Economics, 66*(2), 218-227.

Bateman, I. J., Langford, I. H., Turner, R. K., Willis, K. G., & Garrod, G. D. (1995). Elicitation and truncation effects in contingent valuation studies. *Ecological Economics, 12*(2), 161-179.

Bennett, J., & Adamowicz, V. (2001). Some fundamentals of environmental choice modelling. *The choice modelling approach to environmental valuation*, 37-69.

Bennett, J., & Blamey, R. (2001). *The choice modelling approach to environmental valuation*: Edward Elgar Publishing.

Blamey, R. K., Bennett, J. W., & Morrison, M. D. J. L. E. (1999). Yea-saying in contingent valuation surveys. 126-141.

Boxall, P. C., & Adamowicz, W. L. (2002). Understanding heterogeneous preferences in random utility models: a latent class approach. *Environmental and Resource Economics, 23*(4), 421-446.

Boxall, P. C., Adamowicz, W. L., Swait, J., Williams, M., & Louviere, J. (1996). A comparison of stated preference methods for environmental valuation. *Ecological Economics, 18*(3), 243-253.

Cameron, T. A. (1988). A new paradigm for valuing non-market goods using referendum data: maximum likelihood estimation by censored logistic regression. *Journal of Environmental Economics and Management, 15*(3), 355-379.

Dharmaratne, G. S., Sang, F. Y., & Walling, L. J. (2000). Tourism potentials for financing protected areas. *Annals of Tourism Research, 27*(3), 590-610.

Dümcke, C., & Gnedovsky, M. (2013). The social and economic value of cultural heritage: literature review. *EENC paper*, 1-114.

Eagles, P. F., McCool, S. F., Haynes, C. D., Phillips, A., & Programme, U. N. E. (2002). *Sustainable tourism in protected areas: Guidelines for planning and management* (Vol. 8): IUCN Gland.

Graham, B., Ashworth, G., & Tunbridge, J. (2016). *A geography of heritage: Power, culture and economy*: Routledge.

Green, D., Jacowitz, K. E., Kahneman, D., & McFadden, D. (1998). Referendum contingent valuation, anchoring, and willingness to pay for public goods. *Resource and energy economics, 20*(2), 85-116.

Greene, W. H., & Hensher, D. A. (2003). A latent class model for discrete choice analysis: contrasts with mixed logit. *Transportation Research Part B: Methodological, 37*(8), 681-698.

Grimwade, G., & Carter, B. (2000). Managing small heritage sites with interpretation and community involvement. *International Journal of Heritage Studies, 6*(1), 33-48.

Johnston, R. J., Boyle, K. J., Adamowicz, W., Bennett, J., Brouwer, R., Cameron, T. A., . . . Scarpa, R. (2017). Contemporary guidance for stated preference studies. *Journal of the Association of Environmental and Resource Economists, 4*(2), 319-405.

Lindberg, K. (1998). Economic aspects of ecotourism. *Ecotourism: a guide for planners and managers.*, 87-117.

List, J. A., Sinha, P., & Taylor, M. H. (2006). Using choice experiments to value non-market goods and services: evidence from field experiments. *Advances in economic analysis & policy, 5*(2).

Loomis, J. B. (2014). 2013 WAEA keynote address: Strategies for overcoming hypothetical bias in stated preference surveys. *Journal of Agricultural Resource Economics*, 34-46.

Louviere, J., Hensher, D. A., & Swait, J. D. (2000). *Stated choice methods: analysis and applications*: Cambridge University Press.

Louviere, J. J., Hensher, D. A., & Swait, J. D. (2000). *Stated choice methods: analysis and applications*: Cambridge University Press.

MacDonald, D. H., Ardeshiri, A., Rose, J. M., Russell, B. D., & Connell, S. D. (2015). Valuing coastal water quality: Adelaide, South Australia metropolitan area. *Marine Policy, 52*, 116-124.

Massey, D. (2012). Power-geometry and a progressive sense of place. In *Mapping the futures* (pp. 75-85): Routledge.

McFadden, D. (1986). The choice theory approach to market research. *Marketing science, 5*(4), 275-297.

McIntosh, A. J. (1999). Into the tourist's mind: Understanding the value of the heritage experience. *Journal of Travel & Tourism Marketing, 8*(1), 41-64.

McKercher, B., McKercher, R., & Du Cros, H. (2002). *Cultural tourism: The partnership between tourism and cultural heritage management*: Routledge.

Mitchell, R. C., & Carson, R. T. (2013). *Using surveys to value public goods: the contingent valuation method*: Rff Press.

Moum, T. (1988). Yea-saying and mood-of-the-day effects in self-reported quality of life. *Social Indicators Research, 20*(2), 117-139.

Mowforth, M., & Munt, I. (2015). *Tourism and sustainability: Development, globalisation and new tourism in the third world*: Routledge.

Nuryanti, W. (1996). Heritage and postmodern tourism. *Annals of Tourism Research, 23*(2), 249-260.

Reynisdottir, M., Song, H., & Agrusa, J. (2008). Willingness to pay entrance fees to natural attractions: An Icelandic case study. *Tourism management, 29*(6), 1076-1083.

Rolfe, J., & Bennett, J. (2009). The impact of offering two versus three alternatives in choice modelling experiments. *Ecological Economics, 68*(4), 1140-1148.

Seyedashrafi, B., Ravankhah, M., Weidner, S., & Schmidt, M. (2017). Applying heritage impact assessment to urban development: World heritage property of Masjed-e Jame of Isfahan in Iran. *Sustainable Cities and Society, 31*, 213-224.

Swait, J. (1994). A structural equation model of latent segmentation and product choice for cross-sectional revealed preference choice data. *Journal of retailing and consumer services, 1*(2), 77-89.

Swait, J., & Adamowicz, W. (2001). The influence of task complexity on consumer choice: a latent class model of decision strategy switching. *Journal of Consumer Research, 28*(1), 135-148.

Talboys, G. K. (2016). *Using museums as an educational resource: An introductory handbook for students and teachers*: Routledge.

Tonsor, G. T., & Shupp, R. S. (2011). Cheap talk scripts and online choice experiments: "looking beyond the mean". *American Journal of Agricultural Economics, 93*(4), 1015-1031.

Tuan, T. H., & Navrud, S. (2007). Valuing cultural heritage in developing countries: comparing and pooling contingent valuation and choice modelling estimates. *Environmental and Resource Economics, 38*(1), 51-69.